\documentclass[useAMS,usenatbib,usegraphicx]{mn2e}

\newcommand{\htwo}{H$_2$~}
\newcommand{\Msolar}{$~{\rm M}_{\odot}$~}
\newcommand{\Lsolar}{$~{\rm L}_{\odot}$~}
\newcommand{\echa}{ECHA~J0843.4-7905~}

\title[Molecular Hydrogen emission from disks in the eta Chamaeleontis cluster]{Molecular Hydrogen emission from disks in the eta Chamaeleontis cluster}
\author[S. K. Ramsay Howat and J. S. Greaves]{ S.K. Ramsay
Howat$^{1}$\thanks{Email: skr@roe.ac.uk} and J.S.
Greaves$^{2}$\thanks{ E-mail:
jsg5@st-andrews.ac.uk(JSG)}\\
$^{1}$ UK Astronomy Technology Centre,
Blackford Hill, Edinburgh EH9 3HJ\\
$^{2}$School of Physics and Astronomy, University of St. Andrews,
North Haugh, St Andrews, Fife, KY16 9SS, UK.}
\begin{document}

\date{Accepted; Received; in original form ...}

\pagerange{\pageref{firstpage}--\pageref{lastpage}} \pubyear{2005}

\maketitle

\label{firstpage}

\begin{abstract}

Disks in the 6~Myr old cluster $\eta$ Chamaeleontis were searched for
emission from hot \htwo. Around the M3 star \echa we detect
circumstellar gas orbiting at $\sim$2~AU. If the gas is UV-excited, the
ro-vibrational line traces a hot gas layer supported by a disk of mass
$\sim$0.03\Msolar, similar to the minimum mass solar nebula. Such a gas
reservoir at 6~Myr would promote the formation and inwards migration of
gas giant planets.

\end{abstract}

\begin{keywords} circumstellar matter, infrared:stars, planets:formation,
\end{keywords}

\section{Introduction}

Understanding the longevity of protoplanetary disks around stars, and their
dust and gas content, is an important step in understanding the formation of
planetary systems. The exploration of the dust content of disks has been
carried out from near-infrared to millimetre wavelengths, probing from hot
disks near the stellar surface out to hundreds of AU. The dust mass and
temperature can be deduced from the spectral energy distribution, although
shortwards of the far-infrared the dust emission is optically thick and so
detailed models are needed to understand the distribution of material
\citep{wood02}. Observations of carbon monoxide in disks are used to
estimate the mass of molecular gas, however this is subject to a number of
uncertainties. The CO may be photodissociated; it freezes out on grains and
is only a small amount of the total gas. The H$_2$/CO ratio is highly
variable due to these effects, and CO only effectively traces layers of the
disk above and below the mid-plane. Thus a direct tracer of the bulk of the
gas is to be preferred.

Molecular hydrogen is the principal gas component in disks and so direct
observations of H$_2$ have been undertaken by several groups, using the
rotational-vibrational transitions observable in the near- and
mid-infrared. Mid-infrared observations of the pure rotational lines
offer the most direct measure of the disk mass, as the lines are emitted
by low temperature gas ($\sim$100~K). The line intensity should give a
direct measure of the disk mass for an optically thin disk. To date,
detection of the pure rotational lines has proved difficult. Thi and
co-workers reported detections of copious quantities of \htwo from a
number of different types of source (debris disks, Herbig AeBe stars and
T Tauri's) using ISO SWS (\citealt{thi2001Nature}a,
\citealt{thi2001b}b). However, attempts to replicate these detections in
more sensitive observations with ground-based instruments
(\citealt{richter02}, \citealt{sheret03}) have not detected the
lines. The most likely explanation for this is that the H$_2$ detected
by ISO was elsewhere in the large beam of the spectrometer, rather than
in the circumstellar disk. Both Richter et al. and Sheret et al.
explain their non-detections by pointing out that the disks are
optically thick at mid-infrared wavelengths and that the \htwo may be
therefore be self-shielded. A ground based detection of the
v=0-0 rotational lines from the disk around AB Aurigae has been reported
by \citet{bitner07}, who detect the v=0-0 S(1), S(2) and S(4) lines from
gas at T$\sim$670~K in a disk of mass 0.5\Msolar. Observations with
Spitzer are proving effective for probing the pure rotational lines of
H$_2$ in the circumstellar environment. \citet{lahuis07} have
detected the v=0-0 S(2) and v=0-0 S(3) lines of H$_2$ towards 6 of a
total of 76 T Tauri and Herbig AeBe sources in their sample using the
Spitzer InfraRed Spectrograph. These spectra show the presence of H$_2$ gas at
T$\geq$500~K. 

H$_2$ in disks may also be excited to higher vibration levels and emit
in the near-infrared. The principle excitation mechanisms are by
collisions in warm, dense gas or due to the passage of a shock, by
absorption of a UV photon and subsequent radiative decay or by
X-rays. All of these mechanisms may be in action in a circumstellar
disk. The first detection of emission from the v=1-0 S(1) line at 2.1218$~\mu$m
was been reported by \citet{wkb00} in TW Hya. Following this, the line
has been detected by \citet{bwk03} and \citet{bwk05} for a selection of T Tauri
stars and by by \citet{itoh03} for LkH$\alpha$ 264. In these sources,
the authors conclude that the line emission arises from a quiescent
disk. The line width of the \htwo emission from LkH$\alpha$ 264 suggests
emission from a radius of 1AU from the parent star. For the sources
observed by Bary, Weintraub \& Kastner (2003), the measured widths of the lines
are from $9-14~{\rm km~s}^{-1}$ and the H$_2$ molecules are orbiting from
10--30~AU.  The detection of excited \htwo at these radii offers the
possibility of a large reservoir of cold \htwo associated with the zone
of giant planet formation. 

Estimating the total H$_2$ mass is a complex task, as each of the
possible excitation mechanisms may be at work in a particular
disk. Work so far \citep{bwk03} suggests that the v=1-0 S(1) line
is tracing only a thin hot zone at the top surfaces of the gas
disk, and so is not ideal as a bulk mass tracer. However, the
presence of this line is very useful as an indicator that gas is
still present in a disk; as a diagnostic of the orbital radius
(with high resolution spectroscopy); and for comparison to dust
excesses from the inner disk. These data may address the vital
question of whether a gas supply still exists at a few Myr, when
giant planet cores have grown massive enough to begin to accrete
their thick atmospheres.

The discovery of a new young star cluster, $\eta$ Chamaeleontis at a
distance of just 97pc by \citet{mlf99} presents an exciting opportunity
for studying disk formation at close range. The discovery was made in a
deep ROSAT image. \citet{lyo03} found that an unusually high proportion
of the stars have an L-band excess indicative of the presence of a dust
disk. For their derived age of the stars in the cluster (9~Myr), such a
high disk fraction (60~\%$\pm$13~\%) implied that disk lifetimes may be
higher than previously thought. \citet{hll01} measured the disk
fraction in clusters of different ages, finding a linear relationship
between cluster age and disk fraction. From this, the disk fraction
should drop below 60~\% for an age of 2-3~Myr. Recent work on the $\eta$
Chamaeleontis cluster has challenged the idea that it has an unusually
high disk fraction for the age of the cluster. The age has been revised
downwards slightly to 6 (--1, +2)~Myr \citep{ls04} and new L-band data
\citep{hja05} cast doubt on some of the earlier disk excesses. The
\citet{hja05} results reduce the disk fraction to 17~\%$\pm$11~\%, which
is consistent with the decline in disk fraction with age measured by
\citet{hll01}.Longer wavelength observations with the IRAC camera on
SPITZER confirm that two stars from the cluster (RECX11 and \echa) show
excesses in the $K_s$-[3.6] colour, closest to the K-L colour determined
by \citet{lyo04}, but that more dust disks are revealed from the
[3.6]-[4.5] versus [5.8]-[8.0] colour-colour diagram
\citep{megeath05}. In total, \citet{megeath05} find infared excesses
from 6/15 of the stars in $\eta$ Cham.

Motivated by these discoveries and the successes of Bary et al. in
detecting excited H$_2$, we elected to attempt detections of the v=1-0
S(1) line of H$_2$ in the seven $\eta$~Cham stars identified by Lyo et
al. as having infrared excesses using high spectral resolution infrared
spectroscopy. Four of these sources (RECX5, RECX11, RECX9 and \echa) are
identified as having disks by \citet{megeath05}. The presence of
significant quantities of gas in these 5--8~Myr-old disks would extend
the possible timescales for the formation of gas giant planets.

We report here on the search for H$_2$
in these disks and discuss the connection of gas to the dust content,
excitation mechanisms and age of the stars.

\section[]{Observations and Data Reduction}

The observations were carried out on 25 February 2005 using the
Phoenix high resolution infrared spectrograph on the southern Gemini
telescope at Cerro Pachon, Chile. The observations were carried out in
conditions of thin cirrus; the seeing, as measured from the full width
half maxima of the stars, varied between 0.5~arcsec and 0.8~arcsec
during the night. Phoenix is a single order spectrograph containing a
1024x1024 InSb Aladdin II array \citep{phx03}. With the K4748 filter
and the grating centred at the wavelength of the H$_2$ 1-0 S(1) line
(2.1218~$\mu$m) the wavelength coverage obtained was from 2.11787~$\mu$m to
2.12703~$\mu$m. The slit width selected was 0.35arcsec, matched to four
detector pixels. The spectral resolving power close to the wavelength
of the H$_2$ line was measured to be 62,400, using a telluric OH line
at 2.121774~$\mu$m. This corresponds to a wavelength range per pixel of
0.34$\times10^{-4}~\mu$m. Each source was acquired through the Phoenix
imaging channel. Observations of the science targets were ratioed by the
spectrum of the bright star Beta Car (V=1.68, A2IV) to correct
for atmospheric features. Three observations of the star were made
throughout the night. Wavelength calibration was obtained using the
six brightest telluric absorption features present in the
spectrum. Using this calibration, the wavelengths of the two brightest
OH emission lines present are obtained to better than 0.15$\times10^{-4}~\mu$m. We
take this as a measure of the accuracy of the wavelength
calibration. A flat-field frame was obtained from observations of a
Quartz Halogen lamp in the Gemini Facility Calibration Unit
\citep{gcal}. The flux calibration for the individual observations was
carried out using the known K-band magnitude of the sources
themselves (see Table~1).

The on-chip exposure time for the observations of the targets was
240~seconds, with exposures on the sky obtained by offsetting the source 5~arcsec
along the 14arsec slit. Multiple 240~second exposures were taken and coadded to
produce the total on-source observation time listed in Table 1.

\begin{table*}
 \centering
 \begin{minipage}{140mm}
  \caption{Observational log and target data}
  \begin{tabular}{@{}lccccc@{}}
  \hline
   Name    & RA (J2000)\footnote{Coordinates from \citet{mlf99}.} &  Dec
 (J2000) & Magnitude & Spectral type\footnote{from \citet{ls04}} &
 Total on-source  \\
        &  &  &  &   &observation time   \\
 \hline
 RECX3 & 8 43 37.2  & -79 03 31  & K=9.61 & M3.25 & 48~mins  \\
 RECX5 & 8 42 27.30  & -78 57 48 & K=9.96 & M4 & 48~mins  \\
 RECX6 & 8 42 39   & -78 54 43 & K=9.46 & M3 & 48~mins  \\
 RECX9 & 8 44 16.6 & -78 59 34.4 & K=9.50 & M4.5 & 48~mins \\
 RECX11 & 8 47 01.64  & -78 59 34.4 & K=7.71 & K5.5 & 32~mins  \\
 RECX12 & 8 47 56.77 & -78 54 53.1 & K=8.51 & M3.25 & 32~mins  \\
 ECHAJ0843.3-7905 & 8 41 30.6 & -78 53 07 & K=9.45 & M3.25 & 48~mins  \\
\hline
\end{tabular}
\end{minipage}
\end{table*}

\begin{figure*}
\begin{minipage}{130mm}
  \centering
    \includegraphics[width=120mm,angle=0]{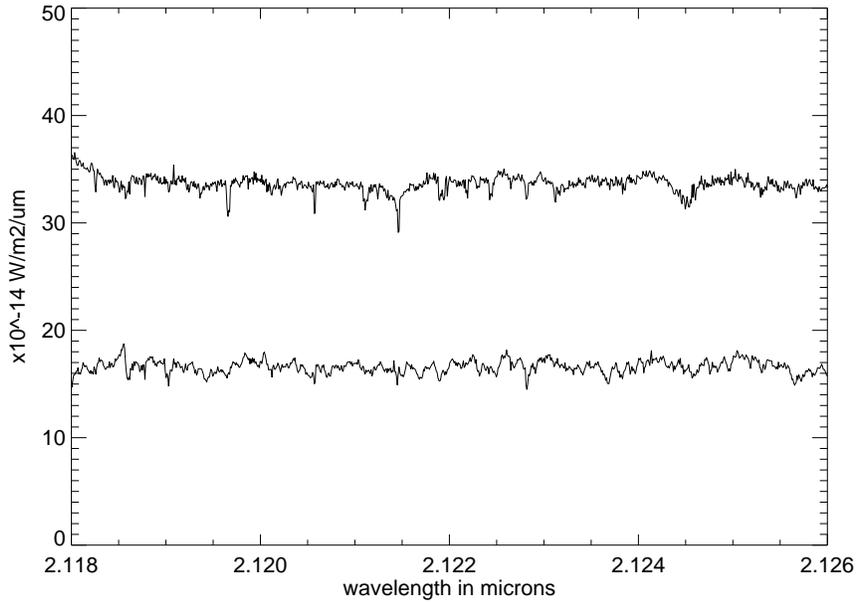}
    \caption{Spectra of RECX11 (upper) and RECX12 (lower). Spectral types are K5.5 and M3.25 respectively.}
\label{spec1}
\end{minipage}
\end{figure*}

\begin{figure*}
\begin{minipage}{130mm}
  \centering
    \includegraphics[width=120mm,angle=0]{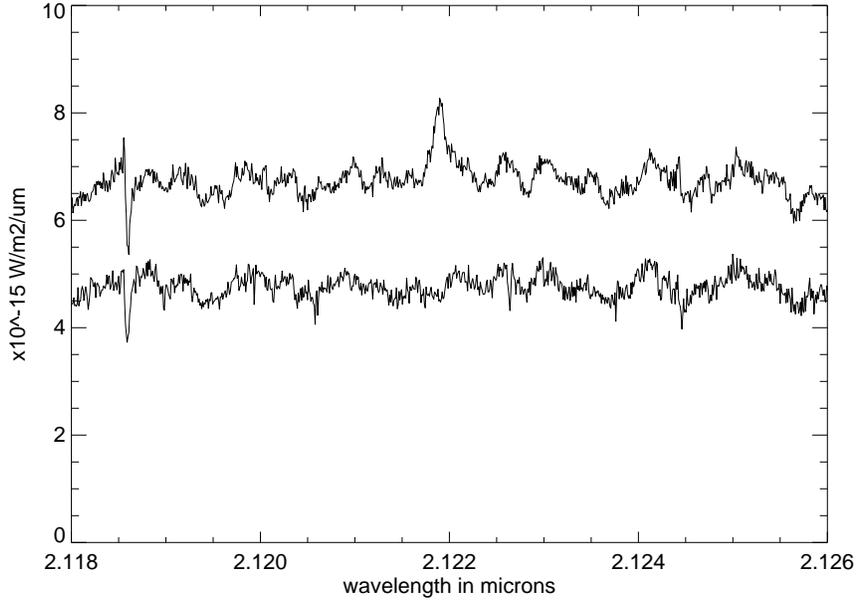}
    \caption{Spectra of RECX6 (lower) and ECHAJ0843.3-7905 (upper). RECX6
    flux is reduced by two to offset it from ECHAJ0843.3-7905. The
    spectral types are M3 and M3.25 respectively.}
\label{spec2}
\end{minipage}
\end{figure*}

\begin{figure*}
\begin{minipage}{130mm}
  \centering
    \includegraphics[width=120mm,angle=0]{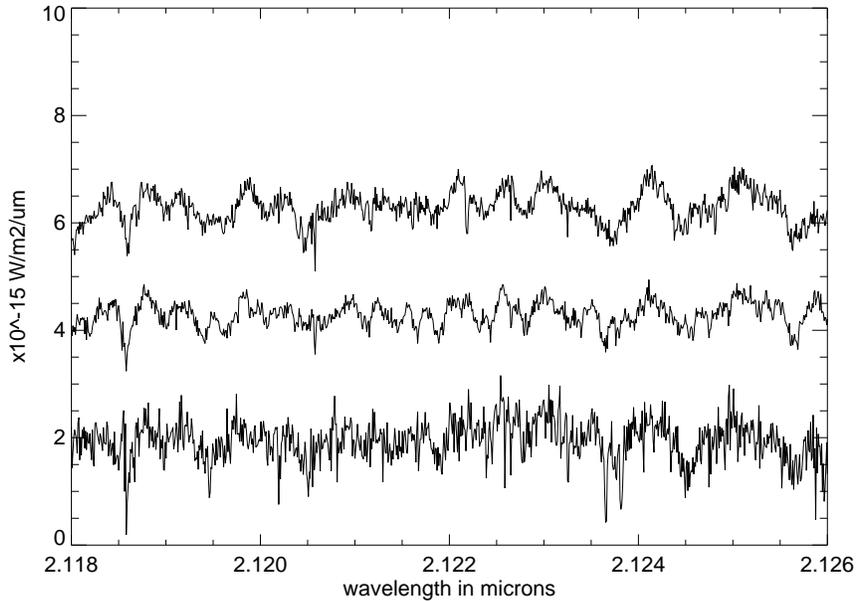}
    \caption{Spectra of RECX3 (lowest), RECX5 (middle) and RECX9
    (lower). RECX3 flux is reduced by four to offset it from RECX9. Spectral types are M3.25, M4 and M4.5 respectively.}
\label{spec3}
\end{minipage}
\end{figure*}

\begin{table*}
 \centering
 \begin{minipage}{140mm}
  \caption{Upper limits and measurements of the H$_2$ line. The
 calculation of the mass of hot H$_2$ and the total disk mass is
 discussed in Section~4.}
  \renewcommand{\thefootnote}{\alph{footnote}}
  \begin{tabular}{@{}lcccccccc@{}}
  \hline
   Name    & Line Flux & Hot H$_2$ mass & Total disk mass \\
        &  $\times10^{-18}$ Wm$^{-2}$  & 10$^{-8}$\Msolar & \Msolar \\
    &&&&&&  \\
 \hline
 RECX3 & $<$1.8&$<$0.36 &$<$0.04--4  \\
 RECX5 & $<$1.2 &$<$0.24 & $<$0.024--2.4 \\
 RECX6 & $<$1.3&$<$0.26 &$<$0.03--3  \\
 RECX9 & $<$1.5 & $<$0.30 &$<$0.03--3 \\
 RECX11 & $<$2.4&$<$0.48 &$<$0.05--5  \\
 RECX12 & $<$2.1&$<$0.42 &$<$0.04--4  \\
 ECHAJ0843.3-7905 & $2.5\pm0.1$ & 0.5 & 0.05--5  \\
GG Tau & $6.9\pm0.5$ & 2.8 & 0.28--28 \\
DoAr21 & $15\pm9$&  8.1 & 0.81--81  \\
LkCa15 & $1.7\pm0.2$& 0.7  & 0.07--7 \\
TW Hya & $1.0\pm0.1$& 0.06 & 6.4$\times 10^{-3}$--0.64  \\
\hline
\end{tabular}
\end{minipage}
\end{table*}

\begin{table*}
 \centering
 \begin{minipage}{140mm}
  \caption{Properties of sample stars and their disks with disk mass
 calculated as discussed in Section~4.}
  \renewcommand{\thefootnote}{\alph{footnote}}
  \begin{tabular}{@{}lcccccccc@{}}
  \hline
   Name    &  Total disk mass  & (K-L) & [3.6]-[4.5] & [5.8]-[8.0] &
 logL$_x$ & cluster or star age\\
        &  \Msolar   &&& &  ergs s$^{-1}$& Myr\\
    &&&&&  \\
 \hline
 RECX3   & $<$0.04--4 &0.14~\footnote{\cite{hja05}} &0.06~\footnote{\citet{megeath05}} &-0.06~\footnotemark[2]& 29.1~\footnote{\citet{mlf99}}& 6\\
 RECX5 & $<$0.024--2.4 & 0.32~\footnotemark[1] & 0.09~\footnotemark[2] &0.48~\footnotemark[2] & 29.0~\footnotemark[3] & 6 \\
 RECX6   &$<$0.03--3 & 0.19~\footnotemark[1]& 0.06~\footnotemark[2] &
 0.0~\footnotemark[2]& 29.5~\footnotemark[3]&  6\\
 RECX9 &$<$0.06--6 & 0.39~\footnotemark[1] &0.19~\footnotemark[2] &0.6~\footnotemark[2]  & 28.4~\footnotemark[3] & 6 \\

 RECX11 &$<$0.1--10 & 0.66~\footnotemark[1] & 0.23~\footnotemark[2]&0.6~\footnotemark[2] &30.1~\footnotemark[3] & 6 \\
 RECX12   & $<$0.04--4 & 0.39~\footnotemark[1]& 0.04~\footnotemark[2]& 0.03~\footnotemark[2]&30.1~\footnotemark[3]& 6\\
 ECHAJ0843.3-7905 & 0.05-5 & 1.32~\footnotemark[1]&  0.47~\footnotemark[2]& 0.91~\footnotemark[2]  & $<$28.5~\footnote{\citet{lcmf02}}  & 6\\
GG Tau & 0.28--28 & 0.87~\footnote{\citet{wg01}} & & & 29.4~\footnote{\citet{bwk03}} & 0.8~\footnote{\citet{siess99}} \\
DoAr21 & 0.81--81 & 0.4~\footnote{\citet{simon95}}& &   & 31.2~\footnotemark[6] & 1.6 \\
LkCa15 & 0.07--7 & 0.57~\footnotemark[2] &&&& 8~\footnotemark[7] \\
TW Hya &6.4$\times 10^{-3}$--0.64 &  0.14~\footnote{\cite{calvet02}} &&& 30.3~\footnotemark[6] & 10 \\
\hline
\end{tabular}
\end{minipage}
\end{table*}

\section[]{Results and Analysis}

Molecular hydrogen has been detected here in one of the seven sources
observed in the $\eta$ Chamaeleontis cluster. The spectra of all the
observed sources are shown in Figures~1--3. The H$_2$ v=1-0 S(1) line is
detected in only ECHAJ0843.3-7905, one of the sources known to have a
circumstellar disk \citep{megeath05}. The spectral types for the stars
range from K5.5 to M4.5 and are listed in Table~1.  The error on the
spectral type is $\pm$0.25 subclasses for the M-stars and $\pm$0.5
subclasses for the K stars \citep{ls04}.

After removal of the telluric features using Beta Car, the observed
stellar continuum from the targets is seen to contain numerous
absorption features that vary with spectral type. To obtain the best
estimate of the continuum level, and hence the optimum measurement of
the H$_2$ line, ECHAJ0843.3-7905 (spectral type M3.25) was ratioed by
the spectrum of stars of the same spectral type (within the errors in
spectral type). There are three such stars: RECX3 (M3.25), RECX6 (M3) and RECX12
(M3.25). The RECX3 spectrum has poor signal/noise and was not used for
the ongoing analysis. The RECX12 and RECX6 spectra were ratioed by Beta Car, as
above, and then normalised to the flux level at $2.1218~\mu$m to provide
a template of the M3 stellar absorption features.

The \echa spectrum was ratioed by the RECX6 and RECX12 templates separately,
providing independent estimates of the \htwo spectrum. Excellent
cancellation of the stellar absorption features is obtained in this way
and comparing the two \echa spectra provides an estimate of the
systematic errors. The spectrum ratioed by the RECX12 template is presented in
Figure~4.

In both spectra, the H$_2$ line is well fitted by a single Gaussian. For
the spectrum ratioed by RECX6 the line width is
$1.40\pm0.08\times10^{-4}~\mu$m centred at
$2.121894\pm0.03\times10^{-4}~\mu$m and, in the spectrum divided by
RECX12, $1.24\pm0.06\times10^{-4}~\mu$m centred at
$2.121903\pm0.02\times10^{-4}~\mu$m. Deconvolving the line widths with
that measured for the unresolved telluric OH lines
($0.35\times10^{-4}~\mu$m) we obtain a line width in the range
$1.35-1.18\times10^{-4}~\mu$m corresponding to a velocity of
$18\pm1.2~{\rm km~s}^{-1}$. The measured flux in the line is
$(2.5\pm0.1)\times10^{-18}{\rm~Wm}^{-2}$.

The presence of \htwo in the cirumstellar environment may be attributed
to molecular gas excited in a disk or in an outflow. For example,
\citet{takami04} concluded that excited \htwo detected in the vicinity of
DG Tauri arises from an outflow within 100AU of the star. This conclusion
is based on a significant blueshift of the line relative to the systemic
velocity of the star and to a spatial offset of the emission from the
star. Moreover, DG Tau is known to have an energetic outflow. Similarly,
\citet{deming04} detected \htwo v=1-0 S(1) emission from a bipolar
outflow associated with KH 15D.

We have considered the possibility that the \htwo emission is from an
outflow associated with \echa. We fit Gaussian functions to the the
spatial profile of the star along the spectrograph slit at the
wavelength of the \htwo line and the adjacent continuum. Subtracting the
continuum from the line emission, and fitting a Gaussian to the results
we find that the peak of the line is centred on the same pixel as the
peak of the continuum (to within 0.5~pixels or 0.04~arcsec) and there is
no evidence for residual \htwo emission in the wings of the line. At the
distance of the $\eta$ Chamaeleontis cluster (97~pc), 0.04~arcsec
corresponds to a scale of $\sim$4~AU. The seeing during the observations
of \echa was 0.7~arcsec (70~AU). Since emission from outflows is observed
at these radii, this is consistent with either an outflow or disk origin
of the \htwo emission. Adopting the radial velocity of the cluster from
\citet{mlf99} for \echa (+16~km~s$^{-1}$) and correcting for the relative
motion of the Earth on the night in question, there is no evidence for
relative motion of the \htwo and the star (to $\sim 1{\rm~km~s}^{-1}$). In addition, the luminosity of the \htwo line (7.2$\times
10^{-7}$\Lsolar) is 20--1000 times fainter than the \htwo luminosity
seen in a range of outflows from Class I and Class II sources
\citep{davis01} or from DG Tau, though we note that there are many
factors, including age of the source, that will affect the luminosity.

We take these diagnostics to mean that it is more likely that the \htwo
emission from \echa arises in a disk and pursue this interpretation in
the following section.

\begin{figure*}
\begin{minipage}{130mm}
  \centering
    \includegraphics[width=120mm,angle=0]{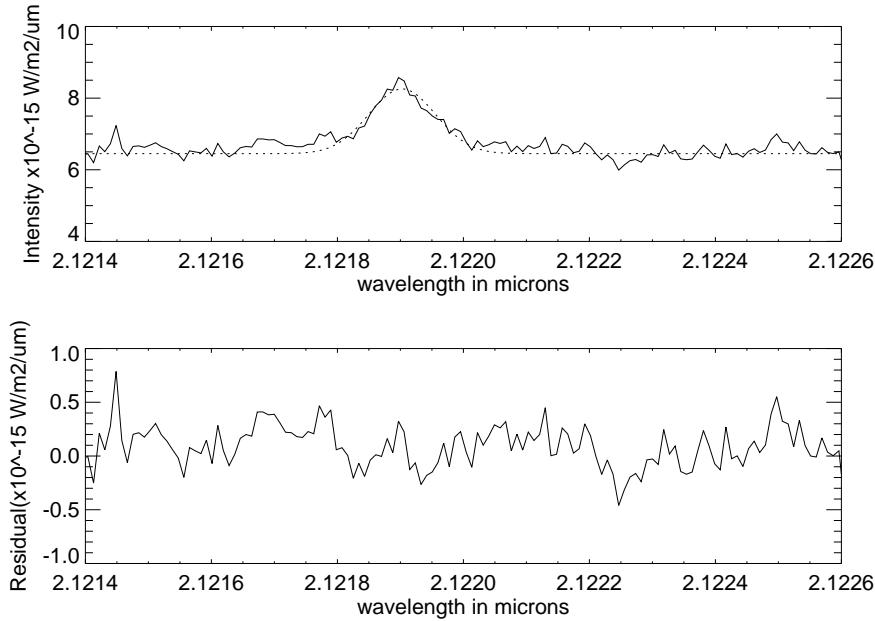}
    \caption{The spectrum of echaJ0843.3-7905 divided by that of RECX12
    with the fit to the H$_2$ 1-0 S(1) line and residuals in the lower
    panel.}
\label{spec_h2}
\end{minipage}
\end{figure*}

\section{Discussion}

Assuming Keplerian rotation around a star of the mass of \echa (taken as
0.4\Msolar), the observed 1-0 S(1) line width implies that the gas is
emitted from a region around a few AU (e.g. 2~AU for a disk inclination
of 45~degrees). The \echa line profile and width is similar to those
observed in the other 4 disk detections, with the exception of LkCa~15
in which a clear double peaked signature is seen \citep{bwk03}. For comparison, the
L-band excesses arise at smaller radii of $\leq 0.1$~AU
\citep{hja05}. The correlation suggests that both trace properties of
the inner disk. The prevalence of single peaked line profiles in \htwo
is perhaps suprising, given that a Keplerian rotation pattern tends to produce
a double-peaked line profile (if the disk is moderately edge-on and
optically thin in the gas line). A line broadening mechanism may be at
work in which case, a narrower Keplerian velocity would imply that the
hot gas is at $>$2~AU. The broader line width may also be due to the
contribution of emission from slower moving gas at larger radii from the
star. \citet{nomura05} show the \htwo emission in the 1-0 S(1) line can
extend over 10~AU with flux greater than 0.1 of the peak flux. Thus a
contribution for an extended disk is consistent with this model and with
a single peaked line.

The v=1-0 S(1) traces regions of hot gas, comprising a small fraction of
the total \htwo reservoir of the disk. Estimating this bulk mass by
extrapolation from the \htwo line strength is difficult, and would
require detailed modelling of the line excitation mechanism(s) to obtain
a robust estimate. We follow previous authors in using the empirical
scaling mechanism derived by \citet{bwk03} to estimate the disk
mass. Calculating the mass of hot H$_2$ from the line strength, we get
0.5$\times10^{-8}$\Msolar for ECHA J0843.3-7905. This calculation
assumes that the excited gas is optically thin, that ro-vibrational levels
of \htwo are populated as for a gas in thermal equilibrium at 1500K and
that the hot \htwo is located at 10~AU to 30~AU. Using the scaling factors
given in \citet{bwk03} gives a mass in the range 0.05 to 5\Msolar for
the \echa disk. The estimated disk masses for our objects are listed
along with those of the other systems with \htwo detections in
Table~2. A disk of the mass implied by the upper range of this scale
would be very suprising for the end of the planet formation timescale
and we consider it to be unfeasible. Indications of the likely disk
masses may be judged from observations by e.g. Rodriguez et al. (2005)
who measure a disk of 0.3-0.4\Msolar for the 0.8\Msolar Class 0 star
IRAS16293-2422B and Andrews \& Williams (2005) who obtain a mass of
0.5\Msolar for the Class I source L1551-IRS5, the latter being the
largest disk observed to date. The assumptions inherent in our
calculation will tend to produce an overestimate of the disk mass. At
radii closer to the star, the gas will be hotter, the mass fraction in
the excited state relatively higher and so the total mass reduced.

Although the mass estimate is very uncertain, important conclusions can
be reached from the detection of gas in this and other systems. A
moderate mass of H$_2$ must still persist around \echa at an age of
6~Myr. It seems very unlikely that a hot component could exist
unsupported by a large reservoir of cool gas in the bulk of the disk
volume. The result for ECHA J0843.3-7905, in conjunction with H$_2$ 1-0
S(1) detections towards LkCa~15 (8~Myr) and TW~Hya (10~Myr), indicates
that a gas reservoir does persist to the ages when gas giant planets are
presumed to form.  For example, in the models of \citet{hbl05}, Jupiter
and Saturn accreted their envelopes after 2--5~Myr. The mass of hot
\htwo against age for all the published \htwo detections is shown in
Figure~5. With the small number of objects so far studied in \htwo, and
the wide range of parameters that affects the observability of excited
\htwo, we can say only that there is a tendency for the older objects to
contain a reduced mass of hot gas. Considering the sources cited, we
also sought correlations between the detection of \htwo line emission
and various other star or disk parameters (listed in Table~3). In
particular, \echa is the star in $\eta$ Cham with the largest excess at
both near and mid-infrared wavelengths, indicating a possible
correlation with the mass of hot gas. However, we find no correlation
between the mass of hot \htwo and L-band excess. Since excitation of the
\htwo may be by UV radation or X-radiation, we also considered these but
again found no correlation. In the following section, we discuss these
two excitation mechanisms.

\begin{figure*}
\begin{minipage}{130mm}
  \centering
    \includegraphics[width=120mm,angle=0]{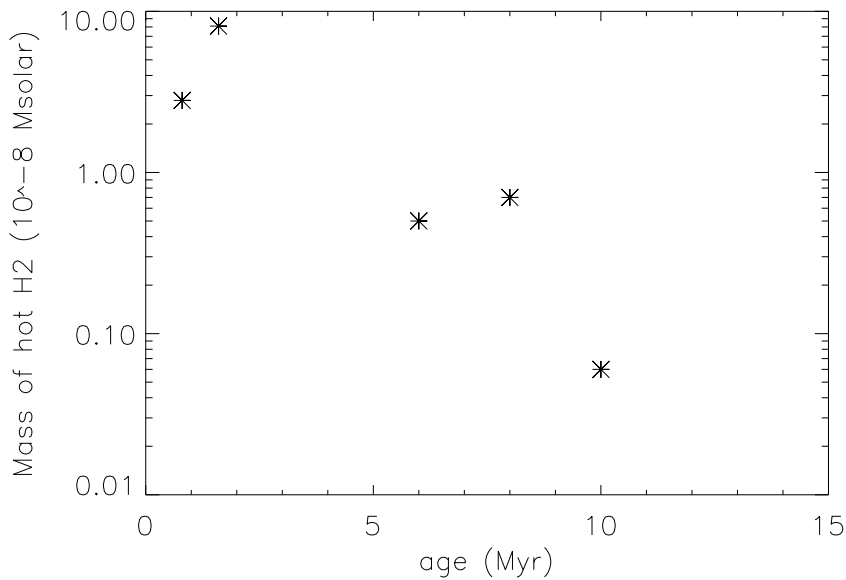}
    \caption{Hot \htwo mass versus age of the source for detections of
\htwo. The \htwo mass for all the sources is calculated using
Equation (2) of \citet{bwk03}.}
\label{mass}
\end{minipage}
\end{figure*}

\subsection{Comparison to models}

Detailed models of H$_2$ excitation in protoplanetary disks have been
published by \citet{nomura05}, \citet{nomura07}. They calculate the gas and dust temperature and
density profiles self-consistently and predict line fluxes for H$_2$ in a
UV-excited scenario, for comparison with the TW~Hya disk. Their model disk
contains 0.006~M$_{\odot}$ of gas and dust and has an outer radius of
100~AU. It may not directly apply to ECHA J0843.3-7905 but some interesting
comparisons may be made.

\citet{nomura05} model the UV and IR emission from \htwo assuming that the
UV radiation field is described either by the blackbody temperature of
the star or, as with many T Tauri stars, that there is an excess of UV
emission characterised by bremsstrahlung emission at higher temperature
than the effective temperature of the star and emission from
Ly~$\alpha$. The latter model is appropriate for TW Hydrae. The `UV
excess' model predicts line strengths of $\sim10^3$ those of the model
without a UV excess. Scaling the predicted 1-0 S(1) flux for TW Hydrae
(at 56~pc) to the distance of \echa (97~pc), the `UV excess' model gives
1.14$\times10^{-18}\rm{W~m}^{-2}$, around five times lower than observed
from \echa. Making the simplifying assumption that the \htwo flux would
scale with disk mass, a disk of 0.03\Msolar would produce the observed
\echa flux, approximately consistent with the lowest mass derived for
the \echa disk by empirical methods. We
therefore consider that the observed flux is consistent with UV
excitation of the 0.03--0.05\Msolar disk by a star with significant UV excess
radiation. The authors were unable to obtain any information on the UV
spectrum of \echa. One anomaly, compared with the model, is that the 1-0
S(1) emission is expected to peak at around 20AU, rather than around
$\sim$2~AU. Since the radial dependence of the observable flux in the
line depends on the disk geometry, we expect the difference to be
attributable to the differences in disk mass.

\citet{mht96} treat the excitation of \htwo due to irradiation by
X-rays. \echa is not detected by ROSAT in the survey of the $\eta$
Chamaelontis cluster by \citet{mlf99} giving an upper limit
of log(L$_x$)=28.5 ergs~s$^{-1}$ \citep{mlf00}. Therefore we can predict only an
upper limit for the 1-0 S(1) emission due to X-ray excitation. Based on
the upper limit for log~(L$_x$), the rate of X-ray
energy deposition at the radius of the emitting gas ($\sim$2~AU) and
assuming an H$_2$ column density of 10$^{22}{\rm~cm}^{-3}$ is 2.5$\times
10^{-23}{\rm ergs~s}^{-1}$. Given a distance to the source of 97pc,
assuming a gas density of 10$^5{\rm~cm}^{-3}$ and an emitting region
stretching from 2~AU to 22~AU, we expect the emission in the 1-0
S(1) to be 2$\times 10^{-19}{\rm~W~m}^{-2}$ \citep{mht96}. Given the
large emitting annulus that we have taken and the X-ray upper limit,
this will most likely be an overestimate and yet is siginificantly
smaller than the observed flux. Thus we conclude that X-ray excitation
is not an important mechanism in this source.

\section{Conclusions}

We have presented a survey of the $\eta$ Chamaelontis cluster in the v=1-0
S(1) line of \htwo. Of four stars surveyed that are currently believed to
have circumstellar disks, \htwo was detected in a single source (\echa). The
line kinematics and flux have been interpreted as arising in a disk,
illuminated by UV radiation from the central star. This result, combined
with those in other clusters, shows that a significant amount of molecular
gas is available for the formation of giant planets on timescales of
6-10~Myr. The gas around \echa is located at $\sim$2~AU, or beyond if the line
width is not all Keplerian. The total mass of gas is estimated at
0.03-0.05\Msolar, which is slightly greater than the Minimum Mass Solar
Nebula (MMSN) of 0.02\Msolar. 

This gas disk meets many of the conditions thought to be necessary for the
formation of giant planets, including long duration, high surface density
and suitable size. For example, the gas reservoir is still substantial at
6~Myr, so a thick gas atmosphere could be added to a Jupiter-analogue core
that has taken $\geq 2$~Myr to form \citep{hbl05}. Further, the surface
density is comparable to the conditions in which Jupiter formed: for a
standard MMSN gas-plus-dust profile of $\Sigma = 1700
r_{AU}^{-3/2}$~g~cm$^{-2}$ \citep{d05}, $\Sigma$(5 AU) is 150~g~cm$^{-2}$,
while for \echa, a gas disk of 0.03-0.05\Msolar exceeds an average $\Sigma =
200$~g~cm$^{-2}$ even if spread over 20~AU radius. The gas also lies at
suitable distances where giant planets should start to form, in particular
in relation to the 'snowline' outside which $\Sigma$ of solid particles is
favourably enhanced because volatiles freeze out into icy mantles on grains.
\citet{lpsc06} show that the snowline in an MMSN disk lies at around 2~AU,
and further in for a low stellar accretion rate such as the $10^{-9}$\Msolar
per year of \echa \citep{llm04}. The Keplerian estimate shows that the \echa
disk extends to at least 2~AU, and so beyond the snowline. Finally, in these
conditions the Type~I migration rate for an Earth-mass proto-planet core is
long: the model of \citet{mg04} suggest a migration timescale $> 2 \times
10^7$ years, so a planet in the \echa disk is unlikely to fall into the 
star. Thus this star is a likely site for giant planet formation, but the 
three other stars surveyed without gas detections are less promising; this 
result is roughly in line with observations of main-sequence Sun-like stars, 
where in long-term monitoring programs $\sim 15$~\% host gas giants
\citep{fischer03}. 

\section*{Acknowledgments}

This paper is based on observations obtained at the Gemini Observatory
for proposal GS-2005B-C-15. The Gemini Observatory is operated by the
Association of Universities for Research in Astronomy, Inc., under a
cooperative agreement with the NSF on behalf of the Gemini partnership:
the National Science Foundation (United States), the Particle Physics
and Astronomy Research Council (United Kingdom), the National Research
Council (Canada), CONICYT (Chile), the Australian Research Council
(Australia), CNPq (Brazil) and CONICET (Argentina). Phoenix was designed
and built at the National Optical Astronomical Observatories and is made
available to Gemini as a visiting instrument. The authors are pleased to
acknowledge the excellent support provided during the observing run by
Verne Smith (NOAO), Claudia Winge (Gemini) and Sybil Adams (Gemini). We
are also grateful to the anonymous referee and to David Weintraub for their helpful comments.

\label{lastpage}

\end{document}